\documentclass[twocolumn,showpacs,epsfig,superscriptaddress]{revtex4}

\usepackage{graphicx}
\usepackage{amssymb}
\usepackage{color}
\usepackage{changes}

\newcommand{\figsizeone}{0.48}
\newcommand{\figsizetwo}{0.8}

\begin{document}

\draft
\title{Oscillation death in coupled counter-rotating identical nonlinear oscillators}
\author{Jung-Wan Ryu}
\affiliation{Center for Theoretical Physics of Complex Systems, Institute for Basic Science (IBS), Daejeon 34126, South Korea}
\author{Woo-Sik Son}
\affiliation{National Institute for Mathematical Sciences, Daejeon 34047, South Korea}
\author{Dong-Uk Hwang}
\email{duhwang@nims.re.kr}
\affiliation{National Institute for Mathematical Sciences, Daejeon 34047, South Korea}
\date{\today}

\begin{abstract}
We study oscillatory and oscillation suppressed phases in coupled counter-rotating nonlinear oscillators. We demonstrate the existence of limit cycle, amplitude death, and oscillation death, and also clarify the Hopf, pitchfork, and infinite period bifurcations between them. Especially, the oscillation death is a new type of oscillation suppressions of which the inhomogeneous steady states are neutrally stable. We discuss the robust neutral stability of the oscillation death in non-conservative systems via the anti-PT-symmetric phase transitions at exceptional points in terms of non-Hermitian systems.
\end{abstract}
\maketitle
\narrowtext

\section{Introduction}

Symmetry and symmetry-breaking transition are very important in the study on the collective behaviors appearing in coupled nonlinear oscillators \cite{Pik01, Kos13}. For example, coupled identical oscillators exhibit complete synchronization resulting from symmetry of the system. In this case, the transition between synchronization and desynchronization is a spontaneous symmetry breaking transition. As a result, the symmetry of the system and corresponding spontaneous symmetry breaking can raise many emergent collective phenomena in coupled nonlinear oscillators.

The study on coupled nonlinear oscillators has focused on the coupled co-rotating oscillators, where two oscillators rotate in same directions. Coupled co-rotating nonlinear oscillators have been widely studied theoretically and experimentally \cite{Pik01}. Recently, there have been a few explorations for coupled counter-rotating oscillators, where two oscillators rotate in opposite directions. In the coupled counter-rotating nonlinear oscillators, a mixed synchronization among different variables emerges, in which some variables are synchronized in phase while other variables can be out-of-phase, i.e., antisynchronized \cite{Kim03, Pra10b, Bho11, Sha12}. The coupled counter-rotating oscillators exist in various systems such as fluid dynamics \cite{Mur02}, biological systems \cite{Tak08}, and dynamical systems \cite{Czo12}.

On the other hand, parity-time (PT) symmetry which originally developed in the context of quantum mechanics supports real eigenenergies of non-Hermitian Hamiltonians describing non-conservative systems \cite{Ben98, Ben07}.  One of the interesting phenomena related to the PT-symmetry is the phase transition between unbroken and broken PT-symmetric phases via exceptional point (EP) where two complex eigenvalues and corresponding eigenstates coalesce because of non-Hermitian degeneracy \cite{Kat96, Hei12}. Especially, the unbroken phases have the properties of conservative systems, e.g., real eigenvalues and non-decaying states, though the PT-symmetric systems are non-conservative. This fascinating systems have been explored in a variety of fields such as optical waveguide \cite{ElG07, Guo09, Rue10}, electronic circuits \cite{Sch11}, atomic physics \cite{Jog10}, magnetic metamaterials \cite{Laz13}, and photonic lattices \cite{Mak08, Reg12, Cer16}. As a counterpart of PT-symmetry, anti-PT-symmetry which has properties completely conjugate to those of PT-symmetric systems has been also reported \cite{Ge13, Wu15, Pen16, Yan17}. There is a transition between pure imaginary and complex eigenvalues of corresponding non-Hermitian Hamiltonians of anti-PT-symmetric systems, in contrast to the transition between real and complex conjugate eigenvalues of those of PT-symmetric ones.

Complete suppression of oscillations have been regarded as a fundamental emergent phenomenon in coupled nonlinear oscillators. Oscillation suppression phenomena can be classified into two different types, i.e., amplitude death and oscillation death, according to different underlying mechanism and manifestations \cite{Kos13}. The amplitude death (AD) refers to a situation where oscillations are suppressed because the stability of fixed point of the system is changed from unstable to stable when individual oscillators are coupled \cite{Sax12}. Thus, AD exhibits a homogeneous steady state (HSS). AD can be achieved in a sufficiently large variance of the frequency distribution \cite{Mir90, Erm90}, existence of time delay in the coupling \cite{Ram98, Str98, Ram99, Ata03}, and various couplings such as conjugate \cite{Kar07, Zhu08}, dynamical \cite{Kon03}, nonlinear couplings \cite{Pra03, Pra10}. The other oscillation suppression phenomenon is the oscillation death (OD) which is a result of symmetry breaking of the system through a pitchfork bifurcation of the steady state, by which the homogeneous steady state splits into two additional branches. Thus, OD is manifested as a stabilized inhomogeneous steady state because each individual oscillator follows different branch. Oscillation suppression phenomena can be used as a control mechanism and an efficient regulator of system dynamics \cite{Zha04, Kim05} as well as an interpretation of neuronal dynamics \cite{Cur10}.

Here, we report the oscillation death of which inhomogeneous steady states are neutrally stable in coupled counter-rotating identical nonlinear oscillators with anti-PT-symmetry. The neutral stability, which means that the variables neither converge nor diverge, usually occurs in conservative systems, in which a conservation of energy prohibits convergence or divergence of nearby phase space trajectories. In contrast to spontaneous generation of the neutral stability in conservative nonlinear systems, it is not easy to find the neutral stability in dissipative nonlinear systems because of breaking of time-reversal symmetry. The neutral stability occurs at the critical border between stable and unstable regions in the phase space, where the maximal value of real parts of eigenvalues of Jacobian matrix equal to zero. 
In this work, in coupled counter-rotating identical oscillators, the radial parts of oscillations are attracted to the limit cycle with constant radius but the angular phase parts related to the rotation on the limit cycle can be neutrally stable because of anti-PT-symmetric transition between unbroken phases with non-rotating phases and broken phases with rotating phases. The non-rotating phases cause the oscillation death with neutrally stable steady states in coupled counter-rotating identical oscillators through the anti-PT-symmetric transition induced bifurcation at EP.

This paper is organized as follows. In Section II we introduce the coupled counter-rotating identical nonlinear oscillators and different emergent states, i.e., limit cycles, AD, and OD. Bifurcation and critical behaviors at EP in phase space are discussed in Section III. In Section IV the noise effect on the neutral stability of the OD region are also discussed. In Section V we summarize our results.

\section{Oscillation death in coupled counter-rotating identical nonlinear oscillators}

\begin{figure*}
\begin{center}
\includegraphics[width=\figsizetwo\textwidth]{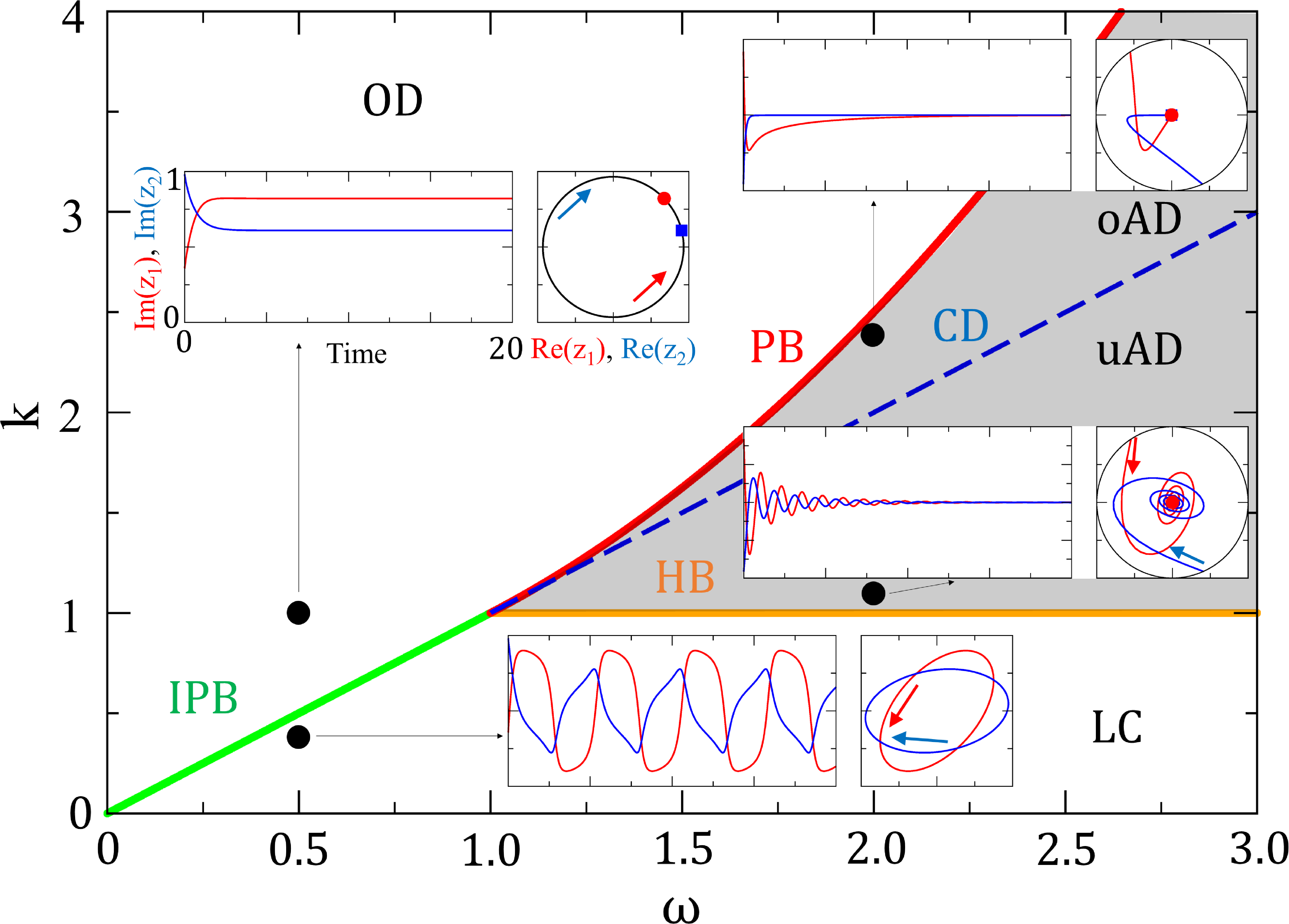}
\caption{(color online) Stability diagram and line of EP (dashed blue and solid green lines) in the parameter space ($\omega$, $k$). LC, uAD, oAD, and OD represent limit cycle, underdamped amplitude death, overdamped amplitude death, and oscillation death, respectively. HB (solid orange line), CD (dashed blue line), PB (solid red line), and IPB (solid green line) represent Hopf bifurcation, critical damping, pitchfork bifurcation, and infinite period bifurcation, respectively. Insets are time series of imaginary parts of variables and corresponding phase portraits. Red and blue trajectories are those of first and second oscillators, respectively. Red circle and blue rectangle are steady states of OD and AD. Rotating directions are represented by red and blue arrows.}
\label{fig1}
\end{center}
\end{figure*}

\begin{figure}
\begin{center}
\includegraphics[width=\figsizeone\textwidth]{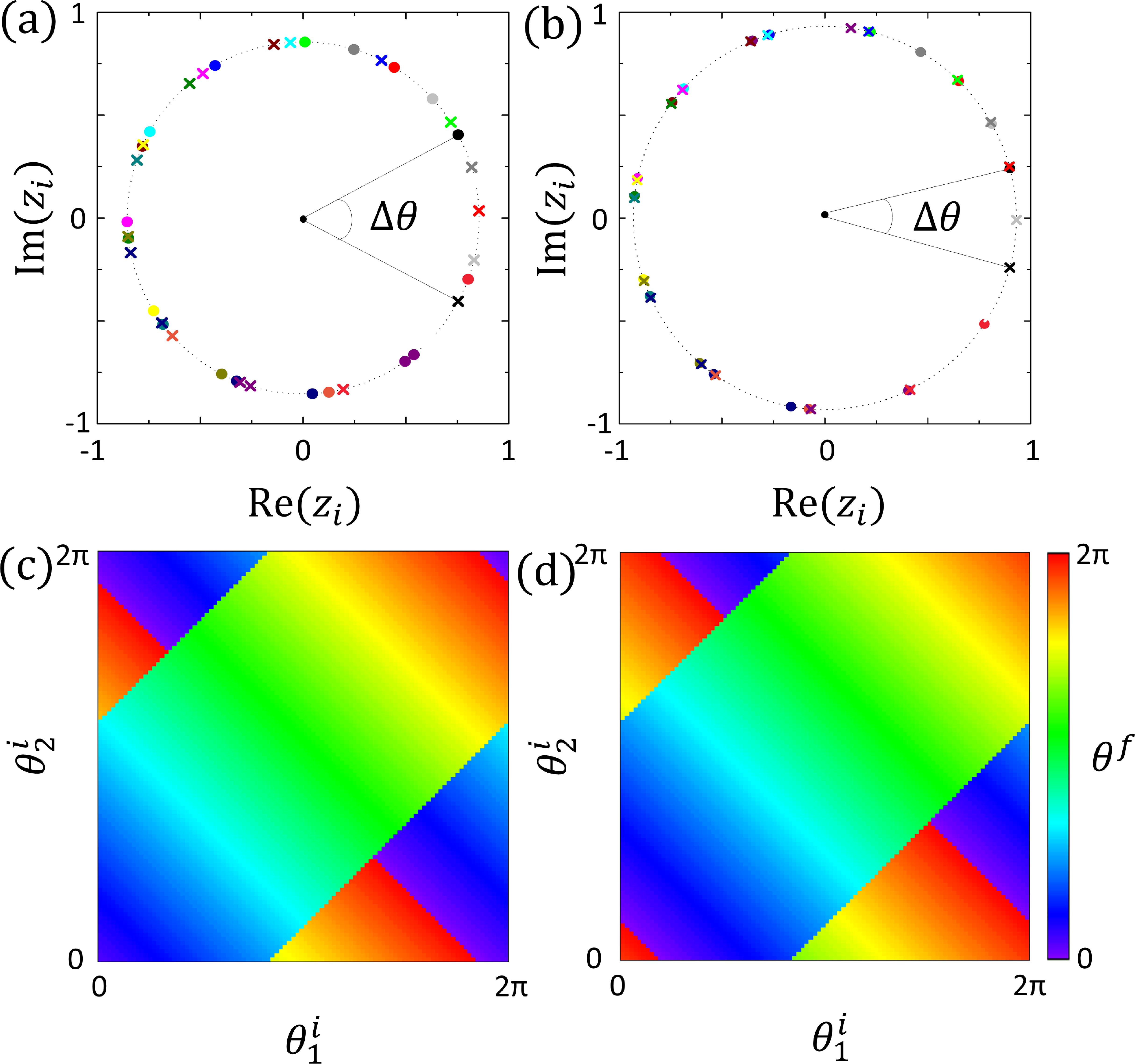}
\caption{(color online) The final steady states of oscillation death when (a) $k=0.6$ and (b) $k=1.0$ with $\omega=0.5$. The circles and crosses denote the steady states of first and second oscillators. The different color represents the different initial points. The angular phases of final states $\theta^f$ of (c) first and (d) second oscillators as a function of the initial angular phases $\theta_{1,2}^i$.}
\label{fig2}
\end{center}
\end{figure}

Let us consider the following system of coupled counter-rotating Stuart-Landau limit cycle oscillators with diffusive coupling,
\begin{eqnarray}
\label{cSL}
 \dot{z}_1=(1 + i \omega - |z_1|^2) z_1 + k (z_{2} - z_{1}), \\\nonumber
 \dot{z}_2=(1 - i \omega - |z_2|^2) z_2 + k (z_{1} - z_{2}),
\end{eqnarray}
where $z_{1,2}$ are complex variables and $k$ is the coupling strength. Here, $\pm \omega$ are the intrinsic angular frequencies of the uncoupled Stuart-Landau oscillators.
The Stuart-Landau limit-cycle oscillator has been widely studied as a paradigmatic model for understanding the collective behaviors such as synchronization and oscillation suppression in coupled nonlinear oscillators because it is a prototypical system exhibiting a Hopf bifurcation and a limit cycle oscillation that can reveal universal features of many practical systems.
In the absence of coupling ($k=0$), two oscillators are attracted to the counter-rotating limit cycles with radii $1$. Coupled nonlinear oscillators with different angular frequencies have been widely studied in terms of synchronization, AD, and OD. The stability diagram for AD is well known in the space ($\Delta w$, $k$) and the critical behavior of AD at EP has been also studied \cite{Erm90, Ryu15}. As shown in Fig.~\ref{fig1}, the stability diagram and line of EP in coupled counter-rotating oscillators of Eq.~(\ref{cSL}) seem to be similar to those in coupled co-rotating oscillators with different angular frequencies, which are well known in the studies on AD, because the diagram shows transition between states as a function of frequency difference and coupling strength. While phase synchronization occurs in coupled chaotic oscillators, if we impose a symmetry, a different type of emergent phenomena, that is, complete synchronization appears in coupled identical chaotic oscillators. Analogously, in this work, by imposing an anti-PT-symmetry on the coupled oscillators, the unprecedented emergent phenomena are observed. The most remarkable phenomenon in coupled counter-rotating identical Stuart-Landau oscillators is the occurrence of OD region which corresponds to the coherent region in the stability diagrams in the precedent studies on AD as shown in Fig.~\ref{fig1} \cite{Erm90}. In addition, the steady states of the OD is neutrally stable, which is totally different from the preceding OD phenomena with inhomogeneous stable steady states. Figure~\ref{fig2} shows the final states of the both oscillators on the limit cycle, starting from different initial points. The angular phases of final states are determined by the angular phases of initial points as shown in Fig.~\ref{fig2}(c) and (d), while the phase difference between final steady states of two oscillators are same without reference to the initial conditions. Although the coupled counter-rotating oscillators are non-conservative systems, the phases of final states memorize the phases of initial points, which is commonly the properties of conservative systems, because of anti-PT-symmetry unbroken phase in the coupled counter-rotating identical oscillators.

In order to reveal the symmetry of the coupled counter-rotating identical oscillators, we use the parity inversion P and time reversal T given by \cite{Ben14}
\begin{eqnarray}
P&:& z_{1,2} \rightarrow z_{2,1}, \\\nonumber
T&:& t \rightarrow -t, ~ i \rightarrow -i.
\end{eqnarray}
Equations~(\ref{cSL}) are not invariant under P and T separately, but they are anti-PT-symmetric which is the anticommutational counterpart of PT-symmetric because $\dot{z}_{1,2} \rightarrow -\dot{z}_{1,2}$ if P and T are applied to the systems simultaneously. Unlike PT-symmetric systems which require balanced gain and loss, anti-PT-symmetric systems keep balanced clockwise and counterclockwise rotations.
We also consider the matrix $M$ obtained from linearized Jacobian Matrix $J$ at the origin, which is given by
\begin{equation}
 \label{jac}
 M = i J = \left(\begin{array}{cc}
 \omega + i (1-k) & i k \\
 i k & - \omega +i (1-k)
\end{array}\right).
\end{equation}
The eigenvalues $\lambda$ of $M$ are complex values because the matrix $M$ is non-Hermitian. Then the time evolution of an eigenvector $\psi$ is given by $\psi(t)=\psi e^{-i \lambda t}$. The real and imaginary parts are the angular frequency of the orbit and the decay (or growing) rates in the vicinity of the origin, respectively. The non-Hermitian matrix $M$ satisfies the anti-PT-symmetry \cite{Pen16, Yan17} and thus there is a spontaneous symmetry breaking of anti-PT-symmetry at a corresponding EP. Non-rotating phases emerge due to the balanced clockwise and counterclockwise rotations in anti-PT-symmetry, while lossless phases occur due to the balanced gain and loss in PT-symmetric systems. As a result, the anti-PT-symmetry of the coupled counter-rotating identical oscillators of Eq.~(\ref{cSL}) is crucial for the OD with neutrally stable steady states.

Rewriting the Eq.~({\ref{cSL}}) using $z_{1,2} = r_{1,2}(t) e^{i \theta_{1,2}(t)}$, we obtain the following equations,
\begin{eqnarray}
\label{cSL2}
 \dot{r}_{1} &=& (1 - {r_{1}}^2) r_{1} - k (r_{1} - r_{2} \cos({\theta_{2} - \theta_{1}})), \\\nonumber
 \dot{\theta}_{1} &=& \omega + k (r_{2} / r_{1}) \sin({\theta_{2} - \theta_{1}}),  \\\nonumber
 \dot{r}_{2} &=& (1 - {r_{2}}^2) r_{2} - k (r_{2} - r_{1} \cos({\theta_{1} - \theta_{2}})), \\\nonumber
 \dot{\theta}_{2} &=& - \omega + k (r_{1} / r_{2}) \sin({\theta_{1} - \theta_{2}}).
\end{eqnarray}
The equation for dynamics of phase difference between two variables, $\Delta \theta = \theta_1 - \theta_2$, can be derived from Eq.~(\ref{cSL2}) as follows,
\begin{eqnarray}
\label{dtheta}
 \dot{\Delta \theta} = 2 \omega - 2 k \sin{\Delta \theta}.
\end{eqnarray}
 Due to the symmetry of two oscillators, $r_1$ equals to $r_2$. When $k<\omega$, there is no fixed point of $\Delta \theta$ since $\dot{\Delta \theta} > 0$. As $k$ increases, stable and unstable fixed points appear via the saddle-node bifurcation at $k=\omega$. If $k>\omega$, $\Delta \theta$ converges into a stable fixed point $\Delta \theta = \arcsin({w/k})$. When $k=0.6$ and $k=1.0$ with $\omega = 0.5$ as shown in Fig.~\ref{fig2}(a) and (b), $\Delta \theta \sim 0.985$ and $\Delta \theta \sim 0.524$, respectively, which accord with the relation $\Delta \theta = \arcsin({w/k})$. Applying the $\Delta \theta =  \arcsin({w/k})$ to the Eq.~({\ref{cSL2}}), $\dot \theta_{1,2} = 0$ is always satisfied. This means two oscillators show oscillation death because two oscillators stop their oscillations at the fixed point but their steady states are not the origins since $r \neq 0$. If we consider the non-symmetric cases of $\omega_1 \neq \omega_2$, where $\omega_{1,2}$ are intrinsic angular frequencies of two oscillators, they exhibit limit cycle oscillations because $\dot \theta_{1,2} \neq 0$. It is noted that the dynamics of $\Delta \theta$ is related to the synchronization of phase oscillators such as Kuramoto model and intermittent route to the synchronization of coupled chaotic oscillators \cite{Pik01}.

\section{Routes to oscillation death}

There exist two routes to OD from limit cycle oscillations as increasing $k$ as shown in Fig.~\ref{fig1}. 

\subsection{Route to OD from limit cycle oscillation via AD}

First when $\omega > 1.0$, the limit cycle oscillation changes into AD via Hopf bifurcation at $k=1$. In this region, AD is an underdamped AD where two oscillators behave like underdamped ones since the fast oscillation is removed due to the relation $\omega_1 + \omega_2 = 0$, where $\omega_{1,2}$ are the intrinsic angular frequencies of the first and second oscillators, respectively. The frequencies of oscillators in underdamped AD regions decrease as $k$ increases till $k = \omega$. At $k=\omega$, the period of oscillations become infinity and the AD is changed into overdamped AD from underdamped AD via critical damping at $k=\omega$. When $k>(1+\omega^2)/2$, the systems show OD after pitchfork bifurcations. Homogeneous stable fixed points at origin for AD are changed into two inhomogeneous neutrally stable fixed points for OD through pitchfork bifurcation in which one stable fixed point changes into one unstable fixed point and two neutrally stable fixed points.

The Hopf and pitchfork bifurcations when $\omega>1$ can be understood by linear stability analysis using the Jacobian matrix $J$ of Eq.~(\ref{jac}) at the origin. The stability diagram of AD in Fig.~\ref{fig1} can be decided by the sign of maximal values of the real parts of complex eigenvalues of the Jacobian Matrix $J$. If the maximal value is smaller than zero, the origin is a stable fixed point and therefore the system exhibits the AD. The shaded regions in Fig.~\ref{fig1} where the maximal value is smaller than zero represent the AD regions. In the regions of LC, uAD, oAD, and OD in Fig.~\ref{fig1}, the Jacobian matrix $J$ has complex conjugate eigenvalues with positive real parts, complex conjugate eigenvalues with negative real parts, real eigenvalues with negative maximum value, and real eigenvalues with positive maximum value, respectively. The Hopf and pitchfork bifurcations can be confirmed by the transitions between two eigenvalues when $k=1$ and $k=(1+\omega^2)/2$, respectively. In addition, the condition for EP, $k=\omega$, can be obtained from double root solutions of complex eigenvalues of the matrix $J$. The EP condition makes critical damping between underdamped and overdamped ADs.

\subsection{Route to OD from limit cycle oscillation without AD}

Next we consider the bifurcation between limit cycle oscillation and OD at EPs when $\omega < 1.0$. So far, it has been known that an EP is a critical point between same phases with same steady states but different transient behaviors such as the critical damping between underdamped and overdamped ADs, not a bifurcation point between two different phases with different steady states, in coupled limit cycle oscillators \cite{Ryu15}. Here, the EP causes the bifurcation between limit cycle and OD. As $k$ increases when $\omega = 0.5$, the period of limit cycle oscillation increases until $k=\omega$ and then the period will be infinity when $k=\omega$ which is the condition of EP. This is an infinite period bifurcation. While the conventional infinite period bifurcation is caused by the saddle-node bifurcation of fixed points on the limit cycle \cite{Str15}, the infinite period bifurcation in coupled counter-rotating identical oscillators occurs due to anti-PT-symmetric transition along with the saddle-node bifurcation of phase difference between two oscillators. In addition, the global bifurcations such as infinite period bifurcation and homoclinic bifurcation can not be explained by the linear analysis near the fixed points because the trajectories in the phase space cannot be confined to a local neighborhood near the fixed point. However, the infinite period bifurcation appearing when $\omega<1$ in anti-PT-symmetric coupled nonlinear oscillators can also be understood by the linearized matrix $M$ at the origin since the dynamics of $\theta$ in Eq.~(\ref{cSL2}) are independent of the dynamics of $r$ when $r_1=r_2$.

\begin{figure}
\begin{center}
\includegraphics[width=\figsizeone\textwidth]{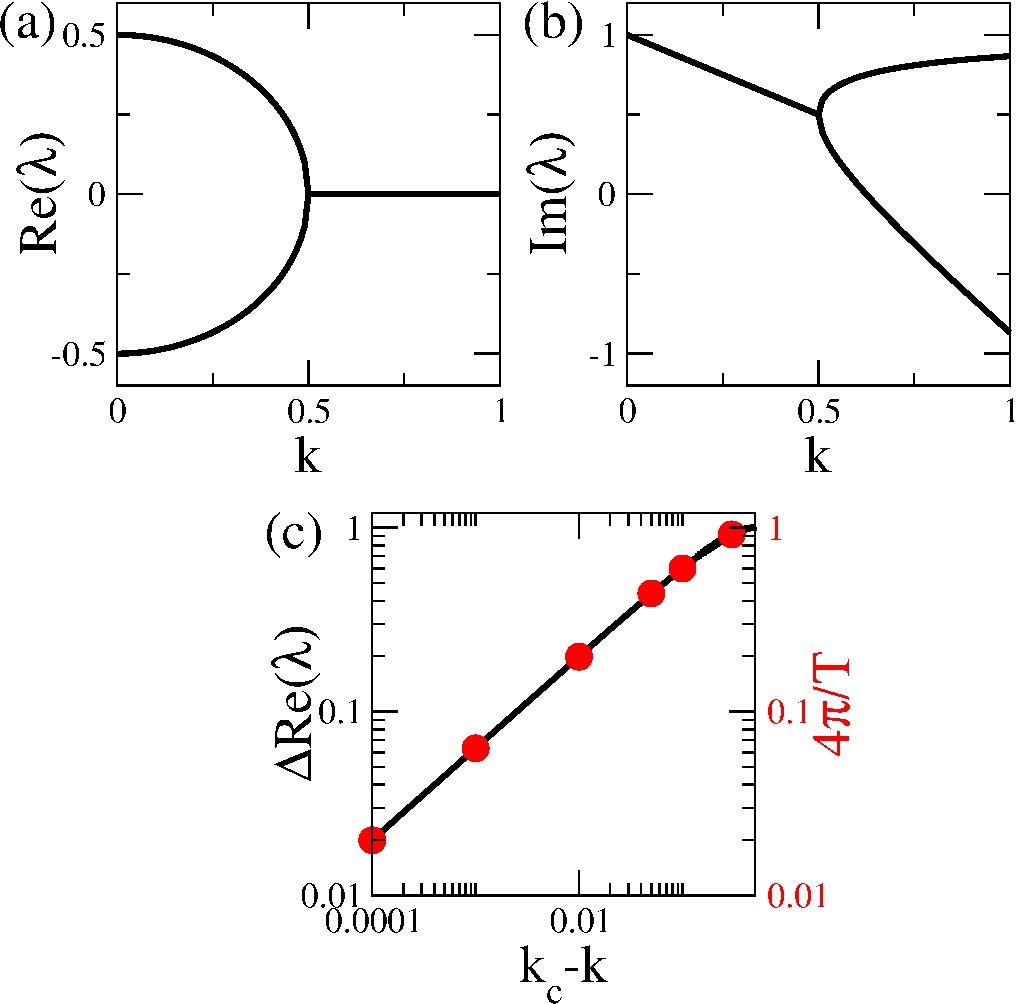}
\caption{(color online) (a) Real and (b) imaginary parts of two eigenvalues of $M$ as a function of $k$ when $\omega=0.5$. EP occurs when $k=0.5$. (c) Difference between two real parts (black line) and periods (red circles) of the oscillations as a function of $k$.
}
\label{fig3}
\end{center}
\end{figure}

Solving the matrix of Eq.~(\ref{jac}), the condition of EP is $k=\omega$, which is the condition of double root of complex eigenvalues of linearized matrix $M$. Figure~\ref{fig3} shows two complex eigenvalues of $M$ as a function of $k$ when $\omega=0.5$ and EP occurs when $k=0.5$, where there is an anti-PT-symmetric transition between broken and unbroken phases. The real parts of eigenvalues are different but the imaginary parts are same when $k<0.5$, of which eigenstates have broken phases. Two different real parts represent the system has two different angular frequencies between which difference determines the angular frequency of the system. The difference decreases and thus the angular frequency (period of oscillation) also decreases (increases) as $k$ increases. The period increases with the relation $( k - k_c )^{-1/2}$ as shown in Fig.~\ref{fig3}(c) because of square root branch property near EP. When $k=0.5$, i.e., the EP, the real parts of eigenvalues are same and the period of oscillation becomes infinite, that is, the infinite period bifurcation occurs at EP. The real parts of eigenvalues equal to zero but the imaginary parts are different when $k>0.5$, of which eigenstates have unbroken phases without oscillations. As a result, an anti-PT-symmetric phase transition occurs at EP obtained from linearized matrix $M$ at the origin, which is related to the infinite period bifurcation in coupled counter-rotating identical nonlinear oscillators.

\section{Noise effect on neutrally stable OD states}

\begin{figure}
\begin{center}
\includegraphics[width=\figsizeone\textwidth]{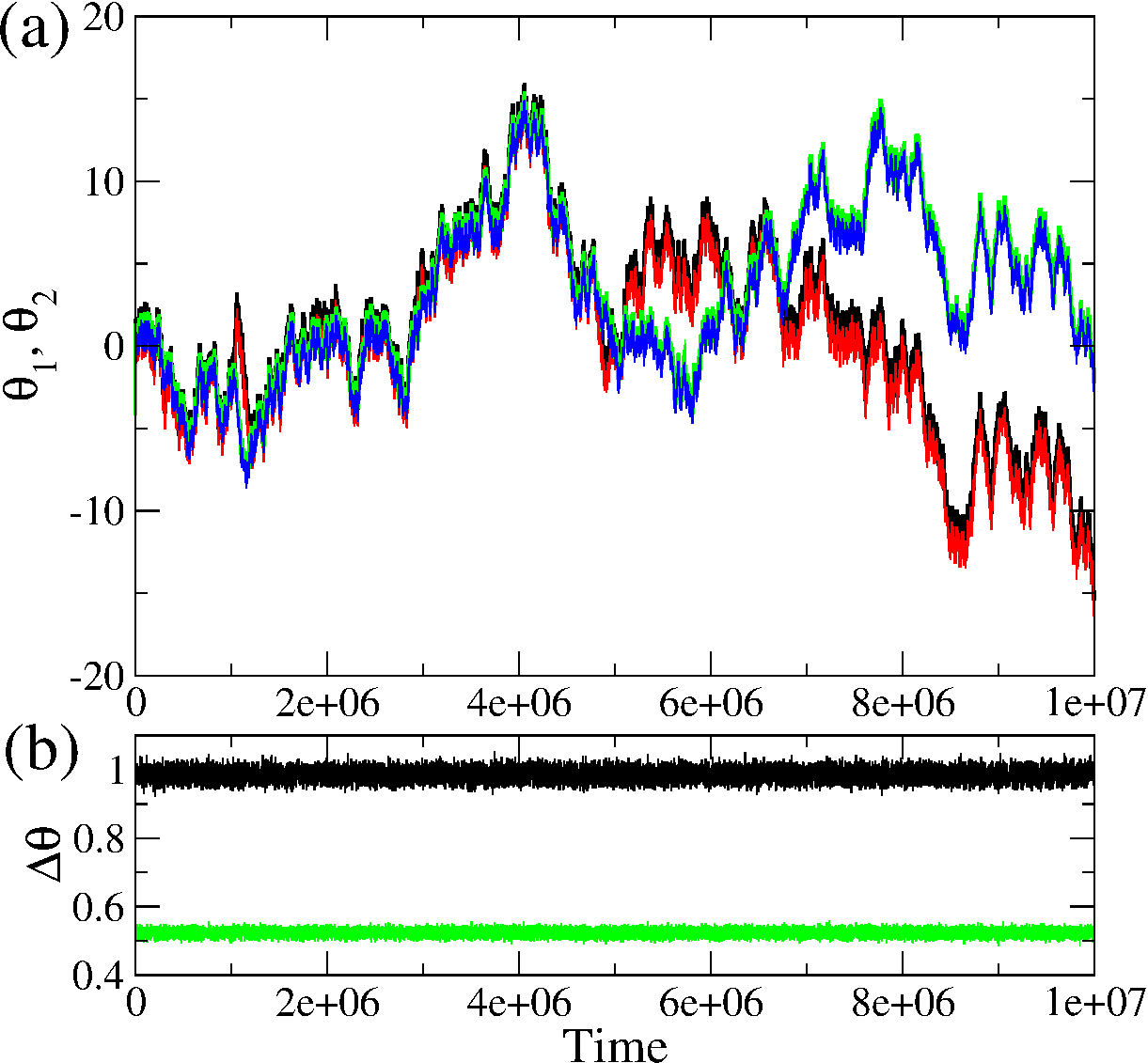}
\caption{(color online) (a) Angular phases of two oscillators when $k=0.6$ and $k=1.0$ with $\omega=0.5$. Black and red trajectories represent angular phases of first and second oscillators, respectively, when $k=0.6$. Green and blue trajectories represent angular phases of first and second oscillators, respectively, when $k=1.0$. (b) The phase difference between two oscillators when $k=0.6$ (black) and $k=1.0$ (green), respectively.
}
\label{fig4}
\end{center}
\end{figure}

Finally, we discuss the noise effect which is unavoidable in natural systems or in experiments on the neutrally stable steady states of OD in coupled counter-rotating identical oscillators. It is well known that stable steady states such as AD or synchronization are robust against presence of noise \cite{Sax12, Tho18}. However, since angular phases of the steady states of the OD in coupled counter-rotating identical oscillators are neutrally stable, the presence of noise leads to the drift of phases. Let us consider the following system of coupled counter-rotating Stuart-Landau limit cycle oscillators with noise,
\begin{eqnarray}
\label{cSL_noise}
 \dot{z}_1=(1 + i \omega - |z_1|^2) z_1 + k (z_{2} - z_{1}) + h \xi_{1}, \\\nonumber
 \dot{z}_2=(1 - i \omega - |z_2|^2) z_2 + k (z_{1} - z_{2}) + h \xi_{2},
\end{eqnarray}
where $\xi_{1,2}$ are complex normal noise and $h$ is noise intensity.

Figure~\ref{fig4} show the angular phases of two oscillators when $k=0.6$ and $k=1.0$ with $\omega = 0.5$ and $h=0.01$. Temporal behaviors of phases show the random walk because they are neutrally stable. Unlike the unbounded nature of random walk behavior of the phases, surprisingly, the phase differences stay around $\Delta \theta \sim 0.985$ and $\Delta \theta \sim 0.524$, respectively, with small fluctuations because they are stable fixed points.

Recently, diffusive systems with anti-PT-symmetry in which the heat transfer in two counter-rotating media have been studied \cite{Li19}. The spontaneous symmetry breaking results in a phase transition from motionless temperature profile as anti-PT-symmetric phase, despite the mechanical motion of the background, to moving temperature profiles as broken phase. Then, their experimental study shows similar observation to our results.

\section{Summary}

We have found the new type of oscillation suppression in coupled counter-rotating identical nonlinear oscillators, of which steady states are neutrally stable. The neutral stability of the oscillation death is originated from the anti-PT-symmetry of the systems. Two routes to the oscillation death have been also studied, one is well-known pitchfork bifurcation from amplitude death and the other is infinite period bifurcation from limit cycle oscillation related to the anti-PT-symmetric phase transtion. The infinite period bifurcation occurs at exceptional point which is the double root solution of the linearized matrix at the origin in the coupled counter-rotating nonlinear oscillators. The infinite periodic bifurcation corresponding to the anti-PT-symmetric transition at EP in coupled counter-rotating nonlinear oscillators is related to the intermittent transition to the anti-synchronization of coupled chaotic oscillators \cite{Kim03}. 

Finally, we expect that new emergent states related to the conservative properties such as neutral stability in dissipative nonlinear systems can be generated by the symmetry recovered by spontaneous symmetry breaking of PT-symmetry such as anti-PT-symmetry of this work.

\section*{Acknowledgments}
We would like to thank H. C. Park for helpful discussion.
This research was supported by Project Code (IBS-R024-D1).
This research was supported by National Institute for Mathematical Sciences (NIMS) funded by the Ministry of Science, ICT \& Future Planning (B19720000).

\end{document}